\documentclass[aps, preprint]{revtex4}
\usepackage{latexsym}
\usepackage{epsfig}

\begin{document}

\title{Effective theory approach to brane world black holes}
\author{Paul L. McFadden}
\email{p.l.mcfadden@damtp.cam.ac.uk}
\author{Neil G. Turok}
\email{n.g.turok@damtp.cam.ac.uk}
\affiliation{D.A.M.T.P., C.M.S., Wilberforce Road, 
  Cambridge, CB3 0WA, UK.}
\date{\today}

\newcommand{\nc}{\newcommand}
\nc{\rnc}{\renewcommand}
\nc{\eg}{\textit{e.g. }}
\nc{\dx}{\mathrm{d} ^4 x}
\nc{\D}{\partial}
\rnc{\d}{\mathrm{d}}
\nc{\tr}{\mathrm{Tr}}
\nc{\gdxdx}{ g_{\mu \nu}(x) \mathrm{d}x^\mu \mathrm{d}x^\nu}
\nc{\ndxdx}{ \eta _{\mu \nu}(x) \mathrm{d}x^\mu \mathrm{d}x^\nu}
\nc{\tgdxdx}{\tilde{g}_{\mu \nu}(x) \mathrm{d}x^\mu \mathrm{d}x^\nu}
\nc{\dxdx}{\mathrm{d}x^\mu \mathrm{d}x^\nu}
\nc{\tz}{\tilde{z}}
\nc{\ty}{\tilde{y}}
\nc{\g}{g_{\mu \nu}}
\nc{\gi}{g^{\mu \nu}}
\nc{\gh}{\hat{g}_{\mu\nu}}
\nc{\psih}{\hat{\psi}}
\nc{\Phih}{\hat{\Phi}}
\nc{\dPhi}{\delta \Phi}
\nc{\dpsi}{\delta\psi}
\nc{\Boxhat}{\hat{\Box}}  
\nc{\gpm}{g^\pm _{\mu \nu}}
\nc{\gp}{g^+ _{\mu \nu}}
\nc{\gm}{g^- _{\mu \nu}}
\nc{\tg}{\tilde{g}_{\mu \nu}}
\rnc{\[}{\begin{equation}}
\rnc{\]}{\end{equation}}
\nc{\bea}{\begin{eqnarray}}
\nc{\eea}{\end{eqnarray}}
\nc{\ie}{\textit{i.e. }}
\nc{\dw}{\mathrm{d}\Omega _2^2}
\rnc{\j}{\mathcal{J}}
\rnc{\tt}{\rightarrow} 
\rnc{\inf}{\infty}
\rnc{\l}{L}
\nc{\RN}{Reissner-Nordstr{\"o}m }
\rnc{\S}{Schr{\"o}dinger }
\nc{\e}{\epsilon}
\nc{\te}{\tilde{\e}}

\begin{abstract}

We derive static spherically-symmetric vacuum solutions
of the low-energy effective action for the two brane Randall-Sundrum model.
The solutions with non-trivial radion belong to a one-parameter family
describing traversable wormholes between the branes and a black hole,
and were first discovered in the context of Einstein gravity with a 
conformally-coupled scalar field.
From a brane world perspective, a distinctive feature of all the
solutions with non-trivial radion is a 
brane intersection about which the bulk geometry is conical but the
induced metrics on the branes are regular.  
Contrary to earlier claims in the literature, we show these solutions are 
stable under monopole perturbations.

\end{abstract}

\maketitle


\section{Introduction}


The brane world provides a fresh perspective on the nature of gravity.
We will consider the simplest possible scenario possessing a warped bulk
geometry, the two brane Randall-Sundrum model \cite{RS1}.
This consists of a pair of four-dimensional positive- and negative-tension $Z_2$ branes 
bounding a five-dimensional bulk with negative cosmological constant.
In addition to the graviton, the spectrum of low-energy gravitational
degrees of freedom includes a massless scalar field, the radion,
pertaining to the separation of the brane pair along the extra dimension.
The low-energy dynamics may then be described by a four-dimensional
effective theory \cite{Ekpyrotic, KS, KSnutshell, Us}, amounting to a specific
scalar-tensor theory of gravity. 

In this article, we derive static spherically-symmetric vacuum 
solutions of the brane world effective theory.  In addition to the well-known 
black string solution with trivial radion \cite{Chamblin}, we find a 
one-parameter family of solutions with non-trivial radion.  These solutions
describe traversable wormholes between the branes, and a black hole. 
For the latter, the geometry induced on the branes is equivalent to that
of the extremal \RN solution, even though there is no electromagnetic
charge present.   
The singularity for this solution is time-like, in contrast to the space-like
singularity of the black string, raising new possibilities for the
fate of infalling matter and the endstate of gravitational collapse
on the brane.
The solutions with non-trivial radion are found to possess a brane intersection
at a finite radius, about which the bulk geometry is conical but the induced 
metrics on the branes are nevertheless regular.

One method of solving the brane world effective theory is to re-cast it in the form 
of Einstein gravity with a conformally-coupled scalar field \cite{Us}, for which the
static spherically-symmetric vacuum solutions are already known 
\cite{BBM, Bekenstein1, Bekenstein2, WH_orig, Barcelo}.  However, here we will instead
pursue a more direct method based around a solution-generating symmetry transformation.
We start in Section II by considering the simple case of tensionless 
branes compactified on an $S^1/Z_2$ orbifold, for which an exact
solution of the bulk geometry may be found. 
Then in Section III we proceed to the physically relevant scenario of
branes with their canonical  
Randall-Sundrum tensions.  Finally, in Section IV we analyse the stability of the
effective theory solutions under monopole 
perturbations.  We show that both the black hole and wormhole
solutions are stable, contrary to earlier 
claims in the literature on Einstein gravity with a conformally-coupled scalar field.

\section{Tensionless branes}

For tensionless vacuum branes compactified on an $S^1/Z_2$ orbifold, the bulk warp is
absent and so the ground state is independent of the fifth dimension $Y$.
Ignoring gauge fields, we introduce the bulk ansatz
\[
\label{bulk_ansatz}
\d s_5^2 = \gdxdx + \Phi^2(x)\d Y^2,
\]
where $x^\mu$, $\mu=0,1,2,3$, parametrise the four conventional
dimensions.  
Inserting this ansatz into the five-dimensional pure
Einstein-Hilbert action and integrating over $Y$, we obtain the
four-dimensional effective action
\[
\label{M_EFT}
S=m_P^2\int \dx \sqrt{-g}\Phi R ,
\]
where $m_P$ denotes the Planck mass.
The corresponding equations of motion are
\bea
\label{M_eom}
\Phi R_{\mu\nu} &=& \nabla_\mu\nabla_\nu \Phi \\
\label{Box_Phi}
\Box \Phi &=& 0 .
\eea
Since the branes are located at constant $Y$, the metric $\g$
appearing in the effective theory may be identified with 
the induced metric on the branes.

Any five-dimensional Ricci-flat metric provides a solution of the 
four-dimensional effective theory.
For example, starting with the five-dimensional black string, we obtain 
the four-dimensional Schwarzschild solution with constant $\Phi$.
Alternatively, we could take the product of Euclidean Schwarzschild and a
flat time dimension as our five-dimensional metric.  This gives a four-dimensional
effective theory solution with non-trivial radion:
\bea
\label{g_soln}
\d s_4^2 &=&  -\d t^2 +(1-\frac{2m}{R})^{-1}\d R^2+R^2\dw \\
\label{Phi_soln}
\Phi &=& \sqrt{1-\frac{2m}{R}}  .
\eea
At $R=2M$, the size of the fifth dimension shrinks to zero
and the branes intersect.
The bulk geometry about this point is conical, as may be seen by
setting $R=2m+\e^2/8m$ and expanding to leading order.  This gives
\[
\d s_5^2 = -\d t^2 +  4m^2 \dw + \d \e^2  +\frac{\e^2}{16m^2}\d Y^2.
\]
For $2(Y^+ -Y^-)=8\pi m$ the geometry is regular at $R=2m$, however 
in general there is a conical singularity.
Nevertheless, all the curvature components remain finite and a straightforward
analytic continuation to $\e<0$ is feasible.  The bulk geometry then describes
a double cone wherein the fifth dimension collapses down to zero size before 
opening up again on the other side.

From the perspective of the four-dimensional effective theory, this
amounts to an analytic continuation of $\Phi$ to values less than
zero.  Changing coordinates to
$R=m(1+y/2+1/2y)$, (\ref{g_soln}) and (\ref{Phi_soln}) become
\bea
\label{M_metric}
\d s^2 &=& -\d t^2 + \frac{M^2}{y^4}(1+y)^4 (\d y^2 +
y^2 \dw ) \\
\label{M_Phi_soln}
\Phi &=& \frac{1-y}{1+y} ,
\eea
where $M=m/2$ and $y$ takes values in the range $0\le y<\inf$.  The brane
intersection has been mapped to $y=1$.
As $\Phi$ passes through zero, the effective action
(\ref{M_EFT}) changes sign but the equations of motion remain
unchanged. 

\section{Randall-Sundrum Branes}

For Randall-Sundrum branes, the effective action may also be written in 
the form (\ref{M_EFT}) (see Appendix A).  However, due to the presence
of a bulk warp, the induced metrics on the positive-tension (plus) and 
negative-tension (minus) branes are no longer identical.  
Instead, they are given in terms of the effective theory metric $\g$ and the
radion $\Phi$ by
\[
\label{new_g_eqns}
\g^{(1)}=\frac{1}{4}(1+\Phi)^2\g , \ \ \ \ \ \ \ 
\g^{(2)}=\frac{1}{4}(1-\Phi)^2\g ,
\]
where brane one is the plus brane and brane two the minus brane for
$\Phi>0$, and the converse holds for $\Phi<0$.  At a brane intersection, 
$\Phi$ changes sign and a plus brane is continued into a minus brane,
and vice versa.  In this fashion, the direction of the bulk warp is 
preserved.

The action (\ref{M_EFT}) has the symmetry
\[
\g \tt \beta \g, \ \ \  \ \ \ \ \Phi \tt \Phi/\beta,
\]
for any constant $\beta>0$.  In the tensionless case, this just amounts to a 
trivial rescaling of the coordinates.  However, in the Randall-Sundrum case, we see
from (\ref{new_g_eqns}) that this symmetry has a non-trivial effect on the brane metrics 
(provided $\Phi$ is itself non-trivial).
Thus, given any one solution of the effective action (\ref{M_EFT}) with non-trivial $\Phi$, 
we may generate a full one-parameter family of solutions.

Applying this procedure to (\ref{M_metric}) and (\ref{M_Phi_soln}), 
we find the one-parameter family of solutions
\bea
\label{BM_metric}
\d s^2 &=& \beta \left(-\d t^2 + \frac{M^2}{y^4}(1+y)^4 (\d y^2 +
y^2 \dw )\right) \\
\label{BM_Phi_soln}
\Phi &=& \frac{1}{\beta}\left(\frac{1-y}{1+y}\right) .
\eea
The metric on the first brane is 
\[
\label{WH}
\d s^2 = \frac{(1+y_0 y)^2}{1-y_0 ^2}\left( -\frac{\d t^2}{(1+y)^2} + \frac{M^2}{y^4}(1+y)^2(\d y^2 + y^2 \dw)\right) ,
\]
where $y_0 = (\beta-1)/(\beta+1)$, and satisfies $|y_0|<1$.
It is easy to see that the metric on the second brane is identical to that on the first, up to an inversion
of the $y$ coordinate:  this is because the effective theory metric (\ref{BM_metric}) is preserved under $y\tt1/y$, 
whereas from (\ref{BM_Phi_soln}), $\Phi\tt -\Phi$.  Hence from (\ref{new_g_eqns}), a $y$ coordinate
inversion transforms brane one into brane two, and vice versa.

The nature of the solution is determined by the parameter $\beta$:
for $\beta>1$, the conformal factor $(1\pm\Phi)^2/4$ is always greater
than zero, and so the causal structure of the brane metrics is
identical to that of (\ref{M_metric}). 
For $\beta \le 1$ however, the conformal factor possesses zeroes and
so we obtain different causal structures.

\begin{figure}[t]
\begin{center}
\includegraphics[width=4cm, keepaspectratio=]{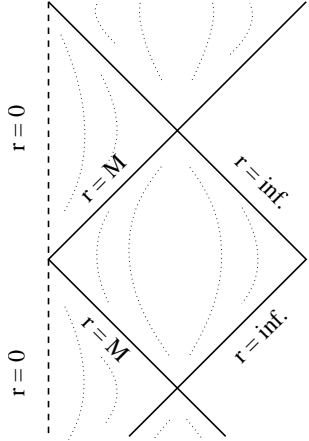}
\end{center}
\caption{Conformal diagram for the BBMB black hole.  Dotted lines
  represent trajectories where $r=\mathrm{const.}$  The vertical
  dashed line represents the central time-like singularity.}
\label{Penrose}
\end{figure}

Explicitly, for $0<\beta<1$ ($y_0<0$), the solution (\ref{WH}) is defined in the range $0<y<1/|y_0|$ and
describes a naked singularity at $y=1/|y_0|$.  As we are
interested in solutions that are regular we will not pursue this case
further.  
For $\beta>1$ ($y_0>0$), the solution is defined in the range $0\le y<\inf$, and
describes a traversable wormhole in which the flat asymptotic region 
at $y=0$ is joined to a second such region at $y\tt\inf$ by an Einstein-Rosen
throat at $y=1$ with no event horizon.
In the case where $\beta=1$ ($y_0=0$), it is useful to substitute $y=M/(r-M)$ giving
\[
\label{BBMB}
\d s^2 = -(1-\frac{M}{r})^2 \d t^2 + (1-\frac{M}{r})^{-2} \d r^2 +
r^2 \dw .
\]
This geometry is equivalent to that of the extremal \RN black hole, even 
though no electric charge is present.  
The Carter-Penrose diagram is given in Figure (\ref{Penrose}).

Historically, the solutions (\ref{WH}) and (\ref{BBMB}) were first discovered in the context
of Einstein gravity with a conformally-coupled scalar field; in particular,
the black hole solution (\ref{BBMB}) is known as the BBMB black hole after its
discoverers Bocharova, Bronnikov, Melnikov \cite{BBM}, and independently
Bekenstein \cite{Bekenstein1, Bekenstein2}.
The traversable wormhole solution was first derived in \cite{WH_orig} and has been discussed 
more recently in \cite{Barcelo}.
That the same solutions re-appear in the present context simply reflects the fact
that it is possible to re-write the brane world effective theory in a different conformal 
gauge (see Appendix A), in which it takes the form of gravity with a conformally-coupled scalar field.
The conformal scalar field is given by
\[ 
\label{psi_soln}
\psi = \sqrt{6}\left(\frac{1-\Phi}{1+\Phi}\right)=
\sqrt{6}\left(\frac{y+y_0}{1+yy_0}\right), 
\] 
and so takes the form
\[
\label{BBMB_psi}
\psi=\sqrt{6}\frac{M}{r-M}
\]
for the BBMB black hole.
However, deriving the solutions of gravity with a conformally-coupled scalar involves
performing a conformal transformation on the solutions of gravity with a \textit{minimally-coupled}
scalar, a more involved calculation than the solution-generating method presented
above.

We now turn to consider the bulk geometry of these solutions.  In the case of branes
with tension, it is not possible to reconstruct the exact bulk geometry 
from a solution of the four-dimensional effective theory alone.  However, 
the behaviour of the radion does provide us with an understanding of
the interbrane separation.  (Specifically, this is given by
$d=2L\tanh^{-1}{|\Phi|}$).  In the
case of the traversable wormhole ($\beta>1$), where $\Phi \tt \pm 1/\beta$ for the flat asymptotics
$y\tt 0$ and $y\tt\inf$, we see that the interbrane separation is asymptotically constant.
In the case of the BBMB black hole ($\beta=1$), the interbrane separation becomes infinite
both at radial infinity ($y=0$) and at the event horizon $r=M$ ($y=1$).
From a brane world perspective, this of course presents no phenomenological difficulties.  Similarly,
in the context of gravity with a conformally-coupled scalar field, Bekenstein has argued that the 
divergence of the scalar field (\ref{BBMB_psi}) at the event horizon does not lead to 
any physical pathologies \cite{Bekenstein2}.

\begin{figure}[t]
\begin{center}
\includegraphics[width=8cm, keepaspectratio=]{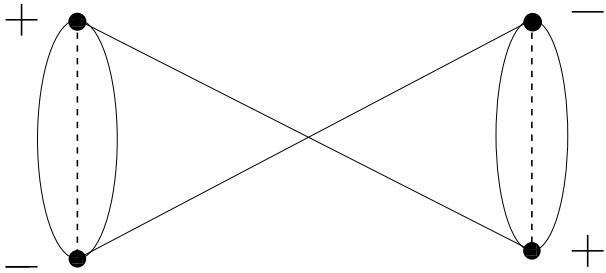}
\end{center}
\caption{The bulk geometry about the brane intersection.  The dashed
  line indicates that the fifth dimension is orbifolded on an
  $S^1/Z_2$.}
\label{dblcone}
\end{figure}

For both the wormhole and the BBMB solution, $\Phi$ vanishes at $y=1$, indicating that
the fifth dimension has shrunk to zero size and the branes are intersecting.
Since the bulk geometry is locally flat in the neighbourhood of any given point, an
approximation to the true bulk geometry about the collision point is provided by neglecting the
brane tensions and using the the Ricci-flat bulk ansatz (\ref{bulk_ansatz}) from the
previous section.  Substituting (\ref{BM_metric}) and
(\ref{BM_Phi_soln}) into (\ref{bulk_ansatz}), and expanding
about $y=1+\e/4M\sqrt{\beta}$, we obtain to leading order 
\[
\d s_5^2 = -\beta \d t^2 +  16M^2\beta \dw + \d \e^2 
+\frac{\e^2}{64M^2\beta^3}\d Y^2.
\]
The bulk geometry about the collision point is again conical (see Figure (\ref{dblcone})), 
but now the conical deficit angle depends on $\beta$ as well as $M$.  
Thus, given a particular asymptotic brane
separation at spatial infinity (fixing $\beta$), there is one specific value of the mass $M$
that removes the conical singularity in the bulk geometry.

Since the brane geometry for the BBMB solution (\ref{BBMB}) is equivalent
to that of the extremal \RN solution, it is interesting to note that 
there have already been numerical attempts 
to find the bulk geometry corresponding to a \textit{fully-general}
\RN geometry on the brane \cite{Chamblin2}.
This work was motivated by the observation that the \RN geometry
solves the projected Einstein equations on the brane \cite{Dadhich},
equivalent to the Hamiltonian constraint equations of general
relativity, thus providing suitable initial data for evolution into the bulk.  
However, the four-dimensional effective theory approach considered here
is far more constraining than that of solving the projected Einstein 
equations.   We have found that in fact only the \textit{extremal} \RN 
geometry is feasible, and even then the bulk geometry will possess a conical 
singularity for general values of the mass parameter.
Our approach additionally allows the ``tidal charge'' parameter
of \cite{Dadhich} to be re-interpreted as the conformal scalar charge of
the BBMB black hole (which is equal to its mass \cite{Bekenstein1}).

Finally, we observe that it is possible to obtain black hole solutions for
which the brane separation at the event horizon is finite by 
detuning the brane tensions slightly from their Randall-Sundrum values.
This follows from recent work \cite{Martinez} in which a black
hole solution was found for conformal scalar gravity with a cosmological
constant and a quartic self-interaction term.  
From \cite{Us}, 
we see that this is simply the brane
frame effective theory for the case where both branes carry additional
cosmological constant terms.  For example, if we work in the plus
brane gauge, then the Lagrangian density for a cosmological term on the minus
brane produces a quartic self-interaction term for $\psi^-$  since
\[
\sqrt{-g^-}\Lambda^-=\sqrt{-g^+}(\psi^-)^4{\Lambda^-\over 36}. 
\]
If the cosmological constants satisfy $\Lambda^+=-\Lambda^->0$,  
the resulting field equations were shown to have a static solution
with extremal Reissner-Nordstr{\"o}m-de Sitter geometry and a non-trivial
conformal scalar field.  
In the case that cosmological constants vanish, this solution reduces
as expected to the  
BBMB solution (\ref{BBMB}).  However, for nonzero cosmological
constants, the solution possesses both an inner, an event, and a cosmological
horizon.  The brane separation at the event horizon is now finite,
as it is at the cosmological horizon.  The simple pole of the conformal scalar
field, corresponding to infinite brane separation, 
is then hidden from view between the inner horizon and the event horizon (which
coalesce in the BBMB limit of $\Lambda^\pm\tt0$).

 
\section{Stability}


\subsection{General case}

In this section we analyse the stability of the BBMB and traversable wormhole solutions
under linear monopole perturbations of the metric and the radion.  
Such perturbations are a distinctive feature
of configurations with a scalar field, and moreover, are the most
likely to display any instabilities that may be present.  This is
because the effective potentials for higher-order multipoles generally
contain centrifugal barrier terms.

Starting with the effective action in the form (\ref{M_EFT}), for which the equations
of motion are (\ref{M_eom}) and (\ref{Box_Phi}), we will take (\ref{BM_metric}) and 
(\ref{BM_Phi_soln}) as the background solution for $\g$ and $\Phi$ respectively.
The perturbed metric $\gh$ and scalar field $\Phih$ may then be written
\bea
\gh (y,t) &=& \g (y) + \delta \g (y,t) \\
\Phih (y,t) &=& \Phi (y) + \dPhi (y,t).
\eea
It is convenient to analyse the perturbations in a gauge in which the scalar
field perturbations decouple from those of the metric.  This gauge is
\bea
\delta (\sqrt{-g}g^{yy}) &=& 0 \\
\delta g_{yt} &=& 0,
\eea
where spherical symmetry additionally implies $\delta g_{y\theta} = \delta g_{y\phi}=0$.
A simple calculation suffices to show that this choice of gauge is always permitted.
Perturbing the scalar field equation (\ref{Box_Phi}), we find
\[
\label{pert_scalar_eqn}
\D_\mu (\delta (\sqrt{-g}g^{\mu y})\Phi'
+\sqrt{-g}\gi\D_\nu\dPhi)=0,
\]
where a prime denotes differentiation with respect to $y$.
Yet with our choice of gauge, 
\bea
\delta(\sqrt{-g}g^{\mu y})&=& \delta^\mu_y
g^{yy}\delta\sqrt{-g}-\sqrt{-g}g^{\mu\lambda}g^{yy}\delta g_{\lambda
  y} \nonumber \\
&=& \delta^\mu_y \delta(\sqrt{-g}g^{yy}) \nonumber \\
&=& 0
\eea
and so the scalar field perturbations do indeed decouple from those of the metric.  
In fact, the scalar field perturbations are the sole dynamical degrees of freedom:
once their behaviour has been determined, that of the metric perturbations follows 
automatically as we will see.

To put the perturbed scalar field equation into Schr{\"o}dinger form
we change variables to
\bea
\label{x_eqn}
x &=& M\left(\frac{1}{y}-y-2\ln{y}\right) \\
\dPhi &=& \frac{y}{M}(1+y)^{-2}\chi e^{i\omega t},
\eea
where $\chi$ is a function of $x$.  Then, 
\[
\D_x = - \frac{y^2}{M(1+y)^2}\D_y
\]
and (\ref{pert_scalar_eqn}) assumes the form
\[
(-\D_x^2+V)\chi = \omega^2 \chi ,
\]
where the potential 
\[
V = \frac{2y^3}{M^2 (1+y)^6}.
\]
Since this potential is positive definite over the entire region of interest $0<y<\inf$,
there are no bound states satisfying the boundary conditions 
$\chi\tt0$ as $y\tt 0$ and $y\tt\inf$.  Thus $\omega^2>0$ and there are no 
exponentially-growing unstable modes.   
Since the parameter $\beta$ has disappeared from the perturbed scalar field equation, 
this conclusion holds for both the wormhole and the BBMB solutions.

To analyse the corresponding metric perturbations, let
\bea
\delta g_{tt} &=& a(y) e^{i\omega t}g_{tt} \\
\delta g_{\theta\theta} &=& c(y)e^{i\omega t} g_{\theta\theta}. 
\eea
We then use the equations of motion in the form
\[
\Phi G_{\mu\nu} = (\nabla_\mu\nabla_\nu-\g\Box)\Phi
\]
(by taking the trace and using (\ref{Box_Phi}) 
one may check that this is equivalent to (\ref{M_eom})).
Perturbing the $yt$ equation to linear order, we find
\[
a(y)=\frac{ -(y^2+1)c+y(y^2-1)c'-\beta y(y+1)^2\dPhi'}{y^2-y+1}.
\]
We may then substitute this into the $yy$ equation.  After using
the perturbed scalar field equation, we obtain the following second order
O.D.E. for $c$, sourced by $\dPhi$ and $\dPhi'$:
\bea
0 &=& y^4(y^2-1)(y^2-y+1)c''-2y^3(y^3-3y^2+1)c' \nonumber \\
&&  +(y^2-1)\left(2y^3+M^2\omega^2(y+1)^4(y^2-y+1)\right)c \nonumber
\\
&& 4\beta y^3(y^2-y+1)\dPhi + 6\beta y^4(y^2-1)\dPhi'.
\eea
At $y=1$ the coefficient of $c''$ vanishes leaving us with the
boundary condition $c'=-2\beta\dPhi$ at this point.
Everywhere else, the equation is regular.  We conclude that the metric
perturbations are well-behaved, and that both the BBMB and wormhole solutions are
stable under monopole perturbations.

We note that this result contradicts an earlier result
in the literature \cite{Bronnikov_WH}.  There, the
stability of the wormhole solution is examined in the Einstein frame 
conformal gauge (see \cite{Us}), in which the action takes the form of gravity with a
minimally-coupled scalar field.  However, in this conformal gauge the scalar
field diverges at the brane intersection leading to a singular Schr{\"o}dinger potential
in the perturbed scalar field equation.  To deal with this singular potential
correctly, it is necessary to impose boundary conditions at the singularity (as
we will see in greater detail in the next section).  As this is not done, the analysis
of \cite{Bronnikov_WH} is incomplete.  
In contrast, the analysis presented
above utilises the $\Phi R$ conformal gauge in which the perturbed scalar field
equation is manifestly non-singular.

\subsection{BBMB black hole}

It has likewise been claimed in the literature on gravity with a conformally-coupled
scalar field that the BBMB solution is unstable to monopole perturbations
\cite{Bronnikov_BH}.
As this stands in contradiction to the results of the previous section, we will now
re-examine the stability of the BBMB black hole in the brane frame
conformal gauges (see Appendix A),
in which the action takes the form of gravity with a conformally-coupled scalar field.

The background metric $\g$ for the BBMB solution is given by (\ref{BBMB}) and the
corresponding conformal scalar field by (\ref{BBMB_psi}).  For $r>2M$, the scalar
field takes values in the range $0<\psi<\sqrt{6}$ and so we are in the plus brane
conformal gauge.  Then at $r=2M$, $\psi=\sqrt{6}$ and the branes intersect.  Nonetheless,
there is a smooth continuation to $r<2M$ for which $\psi$ takes values
greater than $\sqrt{6}$, 
indicating that we have matched onto an interior solution in the minus brane conformal gauge.

However, the action in the minus brane frame possesses ghosts, as
the relative sign between
the gravitational and kinetic terms in (\ref{S-}) is incorrect.  
These ghosts make the analysis of perturbations in the region interior
to the brane intersection much more subtle.
We will find that the gauge choice employed in \cite{Bronnikov_BH}
permits an unphysical influx of scalar charge across the
event horizon.
By contrast, if the perturbations are analysed using the decoupled gauge
introduced in the preceding section this problem is avoided.  In the decoupled
gauge, there is no influx of scalar charge across the event horizon,
and the BBMB black is found to be manifestly stable.
(Additionally, use of the decoupled gauge leads to a regular \S potential for the perturbed
scalar field equation, unlike the gauge choice used in
\cite{Bronnikov_BH}).
Such subtleties were not encountered in the previous section as the
$\Phi R$ conformal gauge employed there is automatically free from ghosts. 

Working in the brane frame gauges and dropping the plus or
minus labels, the vacuum equations of motion are given by
\bea
\label{full_eqn}
(\Box-\frac{1}{6}R) \psi &=& 0 \\
\label{conf_eom}
G_{\mu\nu}&=&T_{\mu\nu},
\eea
where the energy-momentum tensor for the conformal scalar field $\psi$
is 
\[
T_{\mu\nu} = \D_\mu \psi \D_\nu\psi - \frac{1}{2}\g
(\D\psi)^2  +\frac{1}{6}[\g\Box
-\nabla_\mu\nabla_\nu+G_{\mu\nu}]\psi^2 .
\]
Taking the trace of (\ref{conf_eom}) we find that $R=0$, 
allowing us to re-arrange (\ref{full_eqn}) and (\ref{conf_eom}) into the form
\bea
\label{boxpsi=0}
\Box \psi &=& 0 \\
\label{conf_E_eqns}
(1-\frac{1}{6}\psi^2)R_{\mu\nu}
&=&\frac{2}{3}\D_\mu\psi\D_\nu\psi-\frac{1}{6}\g (\D\psi)^2 
 -\frac{1}{3}\psi\nabla_\mu\nabla_\nu\psi .
\eea
The vanishing of the left-hand side of (\ref{conf_E_eqns}) when
$\psi=\sqrt{6}$ corresponds to the divergence of the effective
Newton's constant, $G_N\sim (1-\psi^2/6)^{-1}$.  
	
The perturbed metric $\gh$ and scalar field $\psih$ may be written as
\bea
\label{pert_ansatz}
\gh (r,t) &=& \g(r)+\delta \g (r,t) \\ 
\psih (r,t) &=& \psi(r)+\dpsi(r,t).
\eea
We will first choose to describe the perturbations in the decoupled gauge:
\bea
\label{our_gauge}
\delta (\sqrt{-g}g^{rr}) &=& 0 \\
\delta g_{rt} &=& 0 .
\eea
Defining
\[
\label{x_def}
x = \frac{r(r-2M)}{r-M} + 2M\ln{\left(\frac{r}{M}-1\right)},
\]
so that
\[
\D_x = \left(1-\frac{M}{r}\right)^2\D_r
\] 
(recalling that $y=M/(r-M)$ we see that this $x$ is identical to (\ref{x_eqn})),
we find upon perturbing (\ref{boxpsi=0}) that 
\[
\delta\ddot{\psi} = \dpsi_{,xx}+2\frac{r_{,x}}{r}\dpsi_{,x},
\]
where the dots indicate differentiation with respect to $t$.
This may be put in \S form
by writing $\dpsi = (\chi/r)e^{i\omega t}$, where $\chi$ is a function
of $x$, giving
\[
\label{SE}
(-\D ^2 _x + V)\chi = \omega ^2 \chi
\]
where now
\[
\label{V}
V = \frac{2M}{r^3}(1-M/r)^3.
\]
This potential is positive definite over the region of interest
$M<r<\inf$.
Consequently there are no bound states satisfying the necessary
boundary conditions, $\chi\tt0$ for $x\tt\pm\inf$, and hence no
exponentially-growing instabilities. 

The behaviour of the corresponding metric perturbations may then be
determined using the perturbed Einstein equations,
as shown in Appendix B.  It may be checked that they are everywhere
well-behaved, thus completing the analysis.

We now turn to the earlier analysis of \cite{Bronnikov_BH}.  In this work, the
perturbations are studied in the gauge  
\[
\label{B_gauge}
\delta g_{\theta\theta}=0 
\]
instead of (\ref{our_gauge}).  
With this choice, the metric perturbations no longer
decouple from those of the scalar field in the perturbed scalar field
equation.  
Writing
\bea
\label{Bronnikov_a}
\delta g_{tt} &=& a(r) e^{i\omega t}g_{tt} \\
\label{Bronnikov_b}
\delta g_{rr} &=& b(r)e^{i\omega t} g_{rr}, 
\eea
the perturbed scalar field equation is now
\[
\label{scalar_eom_1}
0=2r^4\delta\ddot{\psi}+(r-M)^2\Big(
\sqrt{6}M(a'-b')-2(r-M)(2\dpsi'+(r-M)\dpsi'')\Big).  
\]
The derivatives $a'$ and $b'$ may then be eliminated using the
perturbed Einstein equations, as shown in Appendix C. 
Following \cite{Bronnikov_BH}, we re-cast the result in \S form by setting
\[
\dpsi = e^{i\omega t}\chi\cdot\frac{(u^3+M^3)}{u^2r\sqrt{u^2-M^2}},
\]
where we have used the shorthand $u=r-M$, and
$\chi$ is a function of $x$ (defined in (\ref{x_def})). 
This yields a \S equation identical to (\ref{SE}), but with a 
different effective potential defined implicitly through \cite{error}
\[
\label{V_B}
V(x)=\frac{u^4}{r^4}\Big[\frac{2M}{r^2u}-\frac{M^2}{(u^2-M^2)^2}
 -\frac{6M^2r^2}{(u^3+M^3)^2} \Big].
\]
This potential decays as $V \sim 2m|x|^{-3}$ for $x\tt
+\inf$ (spatial infinity) and $x\tt-\inf$ (the event horizon).  
Significantly, the potential possesses a double negative pole located
at $r=2M$: in $x$-coordinates, $V \sim -1/4x^2$ close to $x=0$.
As we have seen, this singularity arose from back-substituting 
(\ref{conf_E_eqns}) to eliminate the metric perturbations from 
the perturbed scalar field equation.  (In the
decoupled gauge this step was not necessary and so there we obtained a regular \S problem).

It is claimed in \cite{Bronnikov_BH} that the
quantum-mechanical boundary-value problem corresponding to the potential (\ref{V_B})
(\ie $\chi\tt0$ as $x\tt \pm \inf$) has a spectrum of eigenvalues unbounded from below,
and hence there exists an infinite number of exponentially-growing instabilities.
However, this is no longer the case once appropriate boundary
conditions have been imposed at the singularity:
consider the general form of the solution about $x=0$, 
\[
\label{log_soln}
\chi (x) \sim A\sqrt{|x|}+B\sqrt{|x|}\ln{|x|},
\]
for arbitrary constants $A$ and $B$.
Since $\dpsi \sim \chi/\sqrt{|x|}$, we must impose the boundary condition $B=0$ at $x=0$,
discarding the logarithmic solution that would otherwise lead to the divergence of $\dpsi$.
We are then free to re-scale $\chi$ so as to set $A=1$ without loss of generality.

\begin{figure}[top]
\begin{center}
\includegraphics[width=8cm, keepaspectratio=]{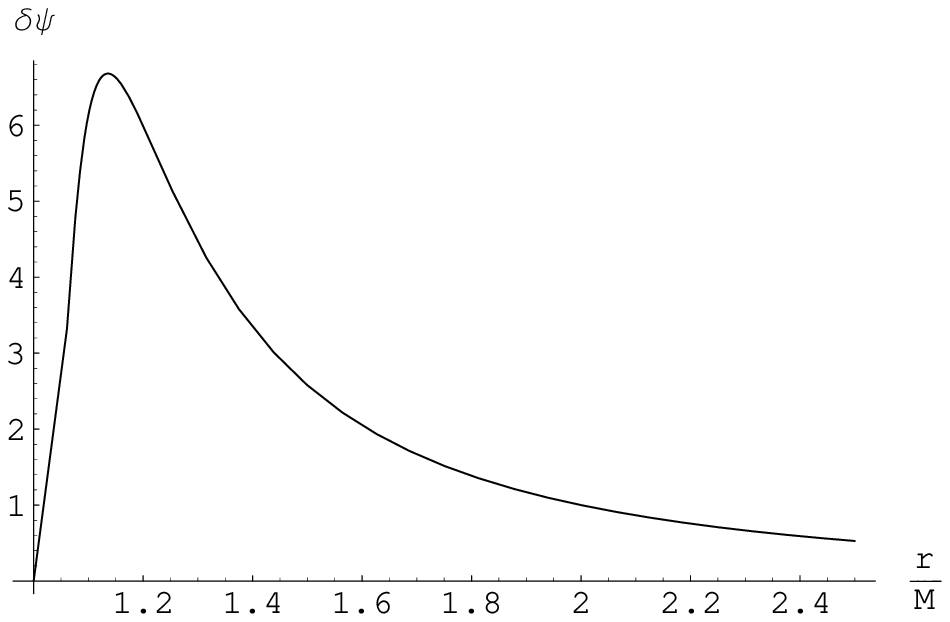}
\includegraphics[width=8cm, keepaspectratio=]{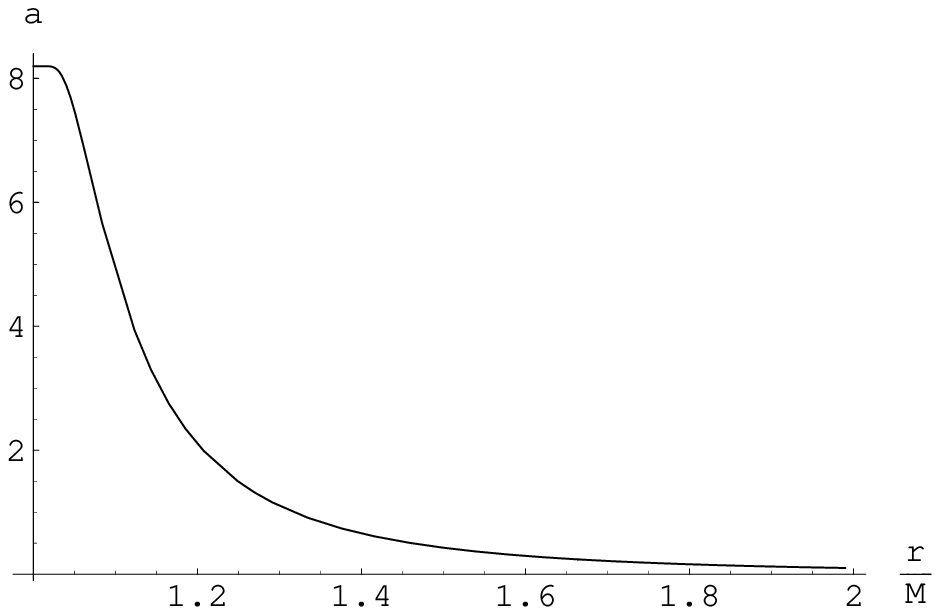}
\includegraphics[width=8cm, keepaspectratio=]{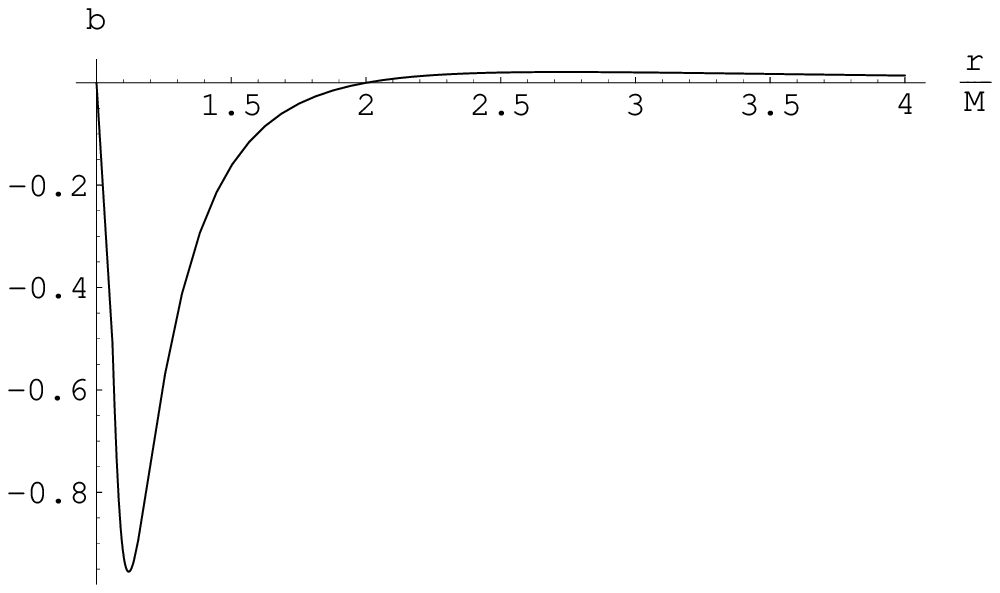}
\end{center}
\caption{Radial profile of the unstable mode with $\Omega = 0.219$.}
\label{mode_fig}
\end{figure}

Using these boundary conditions, we may numerically
solve the perturbed scalar field equation by shooting towards the event horizon
and towards spatial infinity.  Performing this in the original $r$-coordinates
(equation (\ref{r_coord_version}) of Appendix C) for a range of
imaginary frequencies $\Omega=i\omega$ corresponding to an $e^{\Omega t}$ time dependence, 
we may scan the system for unstable modes.
Imposing the additional boundary condition that $\dpsi$ vanish at
spatial infinity we find that there is only 
\textit{one} unstable mode, with frequency $\Omega = 0.219$ 
(see Figure (\ref{mode_fig})). 

For values of $\Omega$ smaller than this the solution always diverges
to negative infinity at the event horizon and at spatial infinity, and for  
values of $\Omega$ greater than this, the solution always diverges to
positive infinity at both boundaries.  (Note the potential (\ref{V_B}) is
even under $x\tt -x$ and for bound states in one dimension there is no
degeneracy, so the eigenmodes must have a definite parity.  Hence a
divergence as $x\tt\inf$ implies a similar divergence for $x\tt-\inf$).

We are then faced with a paradox: in the decoupled gauge (\ref{our_gauge}) the BBMB
solution is manifestly stable, whereas in the alternative gauge
(\ref{B_gauge}) there appears to be one unstable mode satisfying all the
requisite boundary conditions.

To resolve the paradox, we re-interpret the scalar field equation
(\ref{boxpsi=0}) as a conservation law:
\[
\label{cons_law}
\D_\mu \j^\mu = 0,
\]
where the current density
\[
\j^\mu=\sqrt{-\hat{g}}\hat{g}^{\mu\nu}\D_\nu \psih .
\]
By Stokes's theorem, the only way in which the scalar
charge contained in a given spacetime volume can increase is if there is a
corresponding influx of current across the boundary.  Specifically,
consider the spacetime volume bounded by the event horizon and spatial
infinity, and the two arbitrary times $t_1$ and $t_2$.
Applying Stokes's theorem then gives
\[
0=\left[ \int\d\theta\d\phi\d r \j^t\right]^{t_2}_{t_1}
+\left[\int\d\theta\d\phi\d t \j^r\right]^{r=\inf}_{r=M}
\]
as $\j^\theta$ and $\j^\phi$ vanish by spherical symmetry.
Evaluating $\j^t$ we find
\bea
\j^t &=& \sqrt{-g}g^{tt}\delta\dot{\psi}-\sqrt{-g}g^{tt}g^{rr}\delta
         g_{tr}\psi' \nonumber \\
&=& -r^2\sin{\theta}(1-\frac{M}{r})^{-2}\delta\dot{\psi}
\eea
as in both gauges $\delta g_{tr}=0$.  For an instability with 
exponential time dependence $\dpsi=\dpsi(r)e^{\Omega t}$, this becomes
\[
\label{charge}
\left[ \int\d\theta\d\phi\d r \j^t\right]^{t_2}_{t_1} =
-4\pi\Omega\left(e^{\Omega t_2}-e^{\Omega t_1}\right)\int_M^\inf \d r
(1-\frac{M}{r})^{-2}r^2\dpsi(r) . 
\]
Similarly,
\[
\j^r = \sqrt{-g}g^{rr}(\psi'+\dpsi')+\delta (\sqrt{-g}g^{rr})\psi'.
\]
Working first of all in the gauge where $\delta (\sqrt{-g}g^{rr})=0$,
we find that
\[
\left[\int\d\theta\d\phi\d t \j^r\right]^{r=\inf}_{r=M} = 
\frac{4\pi}{\Omega}\left(e^{\Omega t_2}-e^{\Omega
 t_1}\right)\left[(r-M)^2\dpsi'\right]^{r=\inf}_{r=M} ,
\]
which vanishes since $\dpsi'$ tends to zero exponentially rapidly at
both boundaries. 
(The solutions of the \S equation (\ref{SE}) vanish as
$e^{-|\Omega x|}$ as $x\tt\pm\inf$, and here $\dpsi\sim \chi/r$).
Thus, in the decoupled gauge, there is no influx of scalar current across the boundaries.

This is not the case for the gauge choice (\ref{B_gauge}).  
In this gauge, using (\ref{Bronnikov_a}) and
(\ref{Bronnikov_b}) we find 
\[
\left[\int\d\theta\d\phi\d t \j^r\right]^{r=\inf}_{r=M} = 
\frac{2\pi\sqrt{6}M}{\Omega}\left(e^{\Omega t_2}-e^{\Omega
 t_1}\right)\left[b-a\right]^{r=\inf}_{r=M} .
\]
The physical metric perturbations $\delta g_{tt}$ and $\delta g_{rr}$
must vanish at spatial infinity, hence by (\ref{Bronnikov_a}) and
(\ref{Bronnikov_b}) we see that $a$ and $b$ must likewise vanish.
Furthermore, $b$ must also vanish at the event horizon in order for
$\delta g_{rr}$ to remain finite there.  (This also follows from
(\ref{B_b_eqn}), as $\dpsi$ and $\dpsi'$ vanish exponentially at the horizon).
However, $a$ is \textit{not} required to vanish at the horizon: its value there
is determined solely by the radial integral of (\ref{B_a_eqn}).
For the unstable mode discussed above, $a$ is nonzero at the horizon
as may be seen from Figure (\ref{mode_fig}).
Thus, the exponential growth of this mode is supported by the influx
of scalar current across the event horizon.  Equating with
(\ref{charge}), we deduce
\[
a(M)=-\sqrt{\frac{2}{3}}\frac{\Omega^2}{M}\int_M^\inf \d r
(1-\frac{M}{r})^{-2}r^2\dpsi(r) ,
\] 
as is confirmed numerically.

We are now in a position to understand why the two perturbation
analyses performed in different gauges yield conflicting results.  One
gauge permits an influx of scalar current across the horizon; the
other forbids it. Exponentially-growing instabilities are only
possible in the former case.
However, since such fluxes are not physically realistic, we
must instead adopt the latter gauge and conclude that the BBMB black
hole is stable under monopole perturbations.  This conclusion is
supported by our calculations in the $\Phi R$ conformal gauge.


\section{Conclusions}


In this article we have studied a one-parameter family of solutions of
the brane world effective theory corresponding to traversable
wormholes and a black hole. 
While these solutions were already known in the context of gravity with a
conformally-coupled scalar field, our solution-generating method is
novel and our stability analysis corrects that of earlier work.

From a brane world perspective, the most pressing challenge ahead is
to find the exact bulk geometry corresponding to the effective theory
solutions we have found here.  
In addition to knowing the induced
metrics on the branes, a further clue is provided by the conical nature
of the bulk geometry about the brane intersection.
In fact, one possible strategy would be to take the initial data provided by
the brane metrics and solve for the bulk geometry order by order in a
power series expansion about the brane intersection 
(an analogous cosmological example of this is provided in \cite{Tolley}).

Ultimately, we may be led to question the fate of gravitational
collapse on the brane: since the central singularity of the BBMB black hole is
time-like, might it not be possible for matter to avoid the
singularity and instead pass through  
into a second causally-disconnected region of spacetime?

\textit{Acknowledgements:} We thank PPARC for support.


\appendix

\section{Four-dimensional Effective Theory}

In \cite{Us}, the form of the brane world low-energy effective theory was shown
to be highly constrained by symmetry considerations.  
The underlying five-dimensional theory, being pure gravity, is naturally invariant under
diffeomorphisms acting on the bulk.  However, a specific subset of
these bulk diffeomorphisms happen to have the effect of performing a conformal
transformation on the four-dimensional effective theory, whilst preserving the
induced metrics on the branes.
Consequently, the effective theory must itself be conformally invariant.  
This constraint, along with a
knowledge of the static solutions of the theory, is
sufficient to fully determine the form of the effective action up to
two derivatives.  

In the absence of matter, the effective action takes the form 
\bea
\label{EFT}
S &=& m_P^2\int \dx \sqrt{-g}\left(-\psi^+ \Delta \psi^+ + \psi^- \Delta
\psi^-\right) ,
\eea
where $\Delta \equiv \Box - R/6$ and $\psi^\pm(x)$ are two conformally-coupled 
scalar fields related to the positions of the branes in
the extra dimension, $Y^\pm(x)$, by $\psi^\pm=e^{Y^\pm/L}$. 
In this expression $L$ is the bulk anti-de Sitter length 
and $\psi^+>\psi^-$ as $Y^+>Y^-$.
The separation between the branes is then given by $d =Y^+-Y^-= L
\ln {(\psi^+/\psi^-)}$.
The induced metrics on the branes, $\gpm$, are
given in terms of the effective theory metric $\g$ by 
\[
\label{g_eqns}
\gpm = \frac{(\psi^\pm)^2}{6}\g .
\]
As required, the effective action is invariant under conformal
transformations
\[
\label{conf_transf}
\g \tt \Omega^2(x) \g, \ \ \ \ \ \ \psi^\pm \tt \Omega^{-1}(x)\psi^\pm.
\]
The brane metrics $\gpm$ remain invariant by (\ref{g_eqns}).

We will find it useful to fix the conformal gauge in two different ways:
the \textit{$\Phi R$} gauge and the brane frame gauges. 
The $\Phi R$ gauge is obtained by setting $\psi^++\psi^-=\sqrt{6}$ in (\ref{EFT}).  Writing
$\psi^+-\psi^-=\sqrt{6}\Phi$, the effective action in this gauge reads
\[
S = m_P^2\int\dx \sqrt{-g}\Phi R,
\]
and the induced brane metrics are
\[
\label{M_g_eqns}
\g^\pm = \frac{1}{4}(1\pm\Phi)^2\g.
\]
To allow for the possibility of brane collisions, we can generalise the formalism by re-labelling
the plus and minus branes as branes one and two, so that $\Phi=(\psi^{(1)}-\psi^{(2)})/\sqrt{6}$
now takes both positive and negative values.  The brane metrics are then given by (\ref{new_g_eqns}), 
with brane one being the plus brane for $\Phi>0$ and the minus brane for $\Phi<0$.
The interbrane separation is $d=2L\tanh^{-1}{|\Phi|}$.

The brane frame gauges are obtained by setting either of $\psi^+$ or
$\psi^-$ to $\sqrt{6}$, identifying the effective theory metric $\g$ appearing the
action with either the plus or minus brane metric, $\gp$ or $\gm$, respectively.  
In the plus brane gauge, $\psi^+=\sqrt{6}$, the action reads
\[
\label{S+}
S_+=m_P^2 \int\dx\sqrt{-g^+}\left( (1-\frac{1}{6}(\psi^-)^2)R -(\D\psi^-)^2\right),
\]
\ie gravity plus a conformally-coupled scalar field $\psi^-$ taking
values in the range $0\le \psi^-\le \sqrt{6}$. 
In the minus brane gauge, $\psi^-=\sqrt{6}$, the action is
\[
\label{S-}
S_-=-m_P^2\int\dx\sqrt{-g^-}\left( (1-\frac{1}{6}(\psi^+)^2)R -(\D\psi^+)^2\right),
\]
where $\sqrt{6}\le \psi^+<\inf$.  As (\ref{S+}) and (\ref{S-}) are equivalent up to a sign, 
the corresponding vacuum equations of motion are identical; the only difference is
the range of the conformal scalar field.  At a brane collision, where $\psi=\sqrt{6}$,
the solution then smoothly matches from the plus brane gauge to the minus brane gauge 
and vice versa.


\section{Metric Perturbations in Decoupled Gauge}


To analyse the behaviour of the metric perturbations in the decoupled gauge
(\ref{our_gauge})
we must evaluate the perturbed Einstein equations (\ref{conf_E_eqns}) to
linear order.  
This is easily accomplished with the help of a
standard computer algebra package.  Writing
\bea
\delta g_{tt} &=& a(r) e^{i\omega t}g_{tt} \\
\delta g_{\theta\theta} &=& c(r)e^{i\omega t} g_{\theta\theta}, 
\eea
it follows from
the gauge constraints and spherical symmetry that $\delta g_{rr} =
\left(a(r)+2c(r)\right)e^{i\omega t} g_{rr}$, $\delta g_{\phi\phi}
=c(r)e^{i\omega t} 
\delta g_{\phi\phi}$ and all the other
components of $\delta \g$ are zero.
The $tr$ constraint equation may then be solved for $a$:
\bea
\label{a_eqn}
a = -\frac{1}{3r(3M^2-3Mr+r^2)}&&\Big[3r(2M^2-2Mr+r^2)c
  +3r^2(M-r)(r-2M)c' \nonumber \\
&& +\sqrt{6}M(M-r)(M-2r)\dpsi +\sqrt{6}rM(M-r)^2\dpsi'\Big] .
\eea
This equation is everywhere regular.  At the
event horizon, $r=M$, we find $a=c$ and as $r\tt\inf$ we find $a=-c+rc'$.  
Substituting (\ref{a_eqn}) into the $\theta\theta$ Einstein equation then
provides a second order O.D.E. for $c$, sourced by $\dpsi$ and $\dpsi'$:
\bea
\label{c_eqn}
0&=& 3r^3(r-M)^4(r-2M)(r^2-3Mr+3M^2)c'' \nonumber \\
&& -6Mr^2(r-M)^3(r^3-6Mr^2+9M^2r-3M^3)c' \nonumber \\
&& +3r^3(r-2M)[2M(r-M)^3 +\omega^2r^4(r^2-3Mr+3M^2)]c \nonumber \\
&& +6\sqrt{6}M(r-M)^4(r^3-2Mr^2+M^2r-M^3)\dpsi \nonumber \\
&&+ 2\sqrt{6}Mr(r-M)^5(r^2-3M^2)\dpsi' .
\eea
At $r=2M$ the coefficient of $c''$ vanishes and we obtain 
\[
\label{bc1}
2\sqrt{6}Mc'+3\dpsi+2M\dpsi'=0,
\]
thus furnishing a boundary condition for $c'$ at this point.
Elsewhere, the differential equation is regular:
As $r\tt\inf$, (\ref{c_eqn}) reduces to $c''+\omega^2 c=0$ at leading order, 
and at $r=M$ it becomes $c=0$.  
Thus, once the perturbations of the scalar field have been determined,
it is a simple matter to solve for the metric perturbations $a$ and
$c$.


\section{Metric Perturbations in $\delta g_{\theta\theta}=0$ Gauge}


Here we wish to evaluate the metric perturbations in the gauge
(\ref{B_gauge}).  Substituting (\ref{Bronnikov_a}) and
(\ref{Bronnikov_b}) into (\ref{conf_E_eqns}) and expanding to linear order,
the $tr$ constraint equation yields
\[
\label{B_b_eqn}
b=\frac{\sqrt{2}M(r-M)\left(
(M-2r)\dpsi-r(r-M)\dpsi'\right)}{\sqrt{3}r(3M^2-3Mr+r^2)} .
\]
Substituting this into the $\theta\theta$ equation we find
\bea
\label{B_a_eqn}
a' &=& \frac{\sqrt{2}M}{\sqrt{3}r^2(2M-r)(3M^2-3Mr+r^2)^2}\cdot
 \nonumber \\
&& \Big[(-12M^5+21M^4r-42M^3r^2
 +42M^2r^3-20Mr^4+4r^5)\dpsi \nonumber \\
&& -r(r-M)(12M^4+9M^3r -28M^2r^2+18Mr^3-4r^4)\dpsi' \nonumber \\
&& +r^2(r-M)^2(r-2M)(3M^2-3Mr+r^2)\dpsi''\Big] . 
\eea
(The perturbed Einstein equations in this gauge depend only on $a'$,
and not $a$).
Note that in order for $a'$ to be finite at $r=2M$ 
it is necessary that $3\dpsi + 2\dpsi'=0$ at this point.  Eliminating 
$a'$ and $b'$ from the perturbed scalar field equation (\ref{scalar_eom_1}), we find
\[
\label{r_coord_version}
(\mathcal{A}(r)+\omega^2
\mathcal{B}(r))\dpsi+\mathcal{C}(r)\dpsi'+\mathcal{D}(r)\dpsi''=0 ,
\]
where
\bea
\mathcal{A}(r) &=& -6 M^2 (r-M)^2 (-r^3+2Mr^2-M^2r+M^3)  \\
\mathcal{B}(r) &=& r^6 (r-2M) (r^2-3Mr+3M^2) \\
\mathcal{C}(r) &=& 2 r (r-M)^3 (r^4-5Mr^3 +10M^2r^2-6M^3r-3M^4)  \\
\mathcal{D}(r) &=& r^2 (r-M)^4 (r-2M) (r^2-3Mr+3M^2) . 
\eea
This equation is singular when
$\mathcal{D}(r)$, the coefficient of $\dpsi''$, vanishes at $r=2M$.
The general solution about this point is given by the Taylor series 
\[
\label{Taylor_soln}
\dpsi(2M+\e) = \mathcal{P}_1(\e)+j\ln{\e}\cdot\mathcal{P}_2(\e) ,
\]
where $r=2M+\e$ and the polynomials
\bea
\mathcal{P}_1(\e) &=&
k-\left(\frac{j+3k}{2M}\right)\e
+\left(\frac{19j+12k+32(k-j)M^2\omega^2}{8M^2}\right)\e^2 +O(\e^3)  \\
\mathcal{P}_2(\e) &=&
1-\left(\frac{3}{2M}\right)\e+\left(\frac{3+8M^2\omega^2}{2M^2}\right)\e^2
+O(\e^3), 
\eea
for arbitrary constants $j$ and $k$.  Since the single pure logarithmic term
diverges as $\e\tt0$, we must set $j=0$ for $\dpsi$ to be finite. 
(This is the equivalent condition in $r$-coordinates of setting $B=0$
in (\ref{log_soln})). 
The remaining solution branch then
satisfies $3\dpsi+2\dpsi'=0$ at $r=2M$, ensuring that $a'$ is finite.
By re-scaling $\dpsi$ we can without loss of generality set $k=1$,
fixing the form of $\dpsi$ about $r=2M$.  With these boundary conditions
we can then numerically evolve the solution towards infinity and towards
the event horizon.


\end{document}